\documentclass[conference]{IEEEtran}
\IEEEoverridecommandlockouts
\usepackage{cite}
\usepackage{amsmath,amssymb,amsfonts}
\usepackage{algorithmic}
\usepackage{graphicx}
\usepackage{textcomp}
\usepackage{xcolor}
\def\BibTeX{{\rm B\kern-.05em{\sc i\kern-.025em b}\kern-.08em
    T\kern-.1667em\lower.7ex\hbox{E}\kern-.125emX}}

\usepackage{siunitx} % Ensure this package is ok to use
\usepackage{soul}
\usepackage{booktabs}

\newcommand{\fref}[1]{Figure~\ref{#1}}
\newcommand{\eref}[1]{(\ref{#1})}
\newcommand{\sref}[1]{Section~\ref{#1}}
\newcommand{\tref}[1]{Table~\ref{#1}}

\newcommand{\brg}[0]{Biomedical Research Group}

\begin{document}

\title{Audio-based cough counting using independent subspace analysis
\thanks{Identify applicable funding agency here. If none, delete this.}
}

\author{\IEEEauthorblockN{Paul Leamy}
\IEEEauthorblockA{\textit{\brg} \\
\textit{Technological University Dublin\textsuperscript{1}}\\
Dublin, Ireland \\
paul.leamy@tudublin.ie}
% \\
\and

\IEEEauthorblockN{Ted Burke\textsuperscript{1}}
\IEEEauthorblockA{\textit{\brg} \\
% \textit{Technological University Dublin}\\
Dublin, Ireland \\
ted.burke@tudublin.ie}

\and 

\IEEEauthorblockN{Dan Barry}
\IEEEauthorblockA{\textit{QxLab} \\
\textit{University College Dublin}\\
Dublin, Ireland \\
dan.barry@ucd.ie}
% \\
\and

\IEEEauthorblockN{David Dorran\textsuperscript{1}}
\IEEEauthorblockA{\textit{\brg} \\
% \textit{Technological University Dublin}\\
Dublin, Ireland \\
david.dorran@tudublin.ie}
}

\maketitle

% CONFIRM WORD COUNT
\begin{abstract}
% The unprecedented outbreak of novel coronavirus (COVID-19) has seen the number of confirmed cases rise beyond 39 million, with over one million deaths.
% Amongst other health-related issues, a persistent cough is one symptom used to diagnose those who have contracted the virus.
% Early detection of infected individuals has become an urgent priority of health organisations.
% Manually counting coughs is accurate but is time-consuming and requires considerable human effort during a time when medical services are presently under strain.
% In this paper, an algorithm designed to highlight the presence of candidate cough sounds in audio recordings is presented, significantly reducing the time required for manually counting.
In this paper, an algorithm designed to detect characteristic cough events in audio recordings is presented, significantly reducing the time required for manual counting.
% The proposed algorithm uses time-frequency representations and independent subspace analysis (ISA) to detect the presence of sound events that exhibit characteristics of cough sounds, and produces a summary of the recordings that highlights the events detected.
Using time-frequency representations and independent subspace analysis (ISA), sound events that exhibit characteristics of coughs are automatically detected, producing a summary of the events detected.
Using a dataset created from publicly available audio recordings, this algorithm has been tested on a variety of synthesized audio scenarios representative of those likely to be encountered by subjects undergoing an ambulatory cough recording, achieving a true positive rate of 76\% with an average of 2.85 false positives per minute.
% Tested a data from publicly available recordings used to create synthesized audio scenarios (representative of ambulatory cough recording), a true positive rate of 76\% with an average of 2.85 false positives per minute was achieved.

\end{abstract}

\begin{IEEEkeywords}
cough counting, independent subspace analysis, epidemiology, health monitoring
\end{IEEEkeywords}

% ---------------------------------------------------
\section{Introduction} \label{sec:intro}
% Initial spiel
The information contained within the sounds produced by the human vocal tract, such as speech, moaning, sighing, and coughing, present an opportunity to facilitate remote and non-contact monitoring of an individual's health \cite{el2011survey}. 
% In relation to physical health, coughing is a common symptom for which patients seek medical advice \cite{schappert1994national}, especially in the cases of people suffering from coughs in the acute and chronic categories \cite{irwin1998managing}, where a persistent cough can severely impair an individual's quality of life. 
In relation to physical health, coughing is a common symptom for which patients seek medical advice \cite{schappert1994national}, especially in the acute and chronic categories \cite{irwin1998managing}, where a persistent cough can severely impair an individual's quality of life. 
In determining the frequency and severity of a person's cough, a clinician can make a suitable diagnosis relating to a person's cough, and this is the objective of a cough detection/monitoring system \cite{smith2007ambulatory}.
% Having an objective measure of a cough is useful when tracking a disease's progression \cite{french2002evaluation} and this process of early detection and monitoring of a chronic cough becomes even more important when the COVID-19 pandemic is taken into account.
% The World Health Organisation (WHO) stated ``the spread of COVID-19 may be interrupted by early detection, isolation, prompt treatment,...'' \cite{sohrabi2020world}, highlighting the importance of cough monitoring systems to aid with tracking the progress of a disease.
Having an objective measure of a cough is useful when tracking the progression of an illness \cite{french2002evaluation} especially in the process of early detection and monitoring \cite{sohrabi2020world}, highlighting the importance of cough monitoring systems to aid with tracking the progress of a disease.

% Objective assessment of cough recordings has been of interest to clinicians for many years, with cough recording and manual counting recorded as early as the 1960's \cite{woolf1964objective}.
% Although manual cough counting does provide the most accurate method of quantifying cough frequency from audio recordings \cite{turner2014count}, this accuracy depends on how consistent and accurate the person performing the count is, and even self-reporting by patients can be inaccurate too \cite{larson2011accurate,smith2008new}.

% What is shown in this paper
Presented here is an algorithm designed to detect the presence of candidate cough sounds in audio recordings.
The area of cough detection has received notable attention in recent years. 
%  \cite{amoh2016coughDNN,liu2015cough,monge2016effect,monge2018audio,rocha2017detection}, \textbf{...[fill out references here and use most recent for multiple works by one author.]}.
Semi-automatic approaches using hand-crafted features as inputs to probabilistic neural networks \cite{barry2006automatic} and statistical models \cite{matos2006detection,barton2012data} achieved satisfactory results but rely on input from operators to successfully detect cough sounds.
More recent algorithms have used eigenvalue decomposition with random forest classifiers \cite{larson2011accurate}, Hu moments with k-nearest neighbours \cite{monge2018audio}, spectral features with time-delay neural networks \cite{amrulloh2015automatic}, convolutional and recurrent neural networks \cite{amoh2016deep}, and spectral features with support vector machines \cite{you2017cough}.

\begin{figure}[thbp!]
    \centering
    \includegraphics[width=\linewidth]{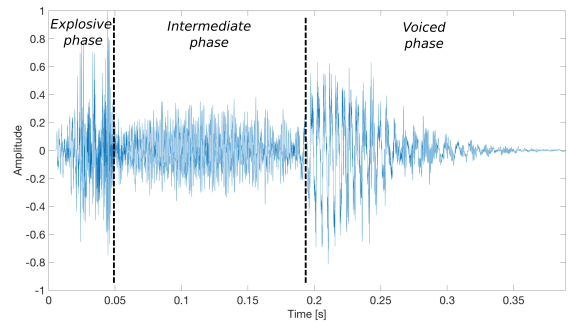}
    \caption{Phases of a cough sound showing explosive phase, intermediate phase, and decaying voiced phase.}
    \label{fig:CoughPhases}
\end{figure}

\begin{figure}[htbp]
    \centering
    \includegraphics[width=\linewidth]{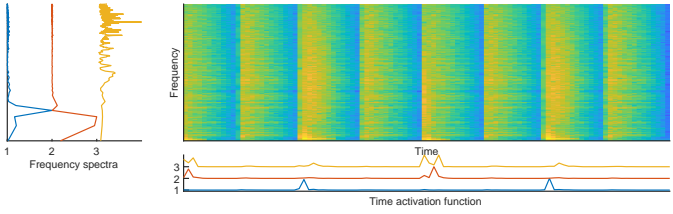}
    \caption{Illustration of ISA being applied to a short drum loop, highlighting the significant time-activation functions and corresponding frequency spectra of each drum.}
    \label{fig:drum_example}
\end{figure}

% The characteristics of cough sounds, as described in \cite{korpavs1996analysis,thorpe2001acoustic,murata1998discrimination}, share characteristics with the signals produced by percussive and harmonic musical instruments.
The nature of cough sounds, as described in \cite{thorpe2001acoustic}, share characteristics with the signals produced by percussive and harmonic musical instruments.
Cough sounds typically begin with a quick onset of wideband noise, stretching across a spectrum up to \SI{20}{\kilo \hertz}, followed by a short stationary period dependent upon the cause of the cough. 
An example of the phases associated with a cough can be seen in \fref{fig:CoughPhases}.
The onset of coughs share similar properties to the percussive characteristics of drums, and in \cite{fitzgerald2002sub} the transient nature of drum sounds were exploited using \textit{independent subspace analysis} (ISA) \cite{casey2000separation} to automatically transcribe drum tracks from audio recordings.
Due to the nature of ambulatory cough recordings, cough sounds produced by a person are likely to be more prominent within these recordings. 
The repeated occurrences of a person's cough and proximity to a microphone means that the cough sound is likely to be one of the sources contributing most variance to the recorded signal. 
Given the variance based nature of ISA, it is expected to produce candidate time-activation functions that coincide with cough events. 
An example of this algorithm's output is illustrated in \fref{fig:drum_example}, which highlights the time-activation functions and corresponding frequency spectra for a short drum loop containing three sounds (kick drum, snare, hi-hat). 
The approach presented here builds on and refines the ISA approach for the purpose of cough event detection.

% ---------------------------------------------------
\section{Methodology} \label{sec:methodology}
Many existing approaches to cough detection are trained in a manner that aims to achieve generalisation such that algorithms can correctly identify cough sounds in a wide variety of recording scenarios.
While generalisation is desirable in detection algorithms, it may not be required to achieve satisfactory results when detecting the presence of repeating events in more constrained scenarios.
In ambulatory cough recordings, the coughs present are likely to come from a single person, which presents an opportunity, namely the characteristics of the coughs to be detected in an audio recording are likely to share similar time-frequency domain features.

\begin{figure}[htbp!]
    \centering
    \includegraphics[width=\linewidth]{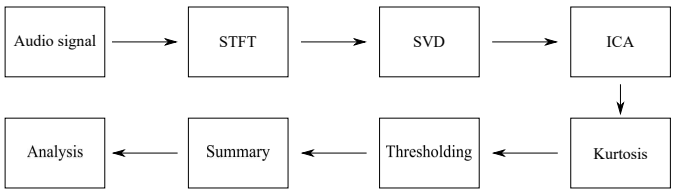}
    \caption{Block diagram illustrating the process of the proposed algorithm. Further detail of each block in this process is presented in \sref{sec:methodology} of this paper.}
    \label{fig:algorithm_overview}
\end{figure}

The algorithm proposed in this paper is designed to summarise an ambulatory recording, producing a number of short audio clips, each of which is likely to contain a cough event. 
Using SVD to analyse the spectrogram of an audio signal, a set of time-activation functions is produced which then undergoes independent component analysis (ICA). Together, SVD followed by ICA is known as independent subspace analysis (ISA).
Significant peaks in these time-activation functions are then used as markers for candidate cough events since these peaks correspond to the presence of high variance events in the audio recording.
Using a suitable threshold, candidate events can be marked and extracted from the input audio signal and presented to the clinician for further analysis and verification.
The algorithm overview is illustrated in \fref{fig:algorithm_overview}, and a \mbox{MATLAB} implementation of the algorithm is available at \cite{leamy_2020_isacough}.

% ----------------------------------------------
\subsection{Short-time Fourier transform} \label{ssec:STFT}
The first task of this algorithm is to compute the complex short-time Fourier transform (STFT) $X$ of the input data $x$. 
This is computed using \eref{eq:STFT}.

\begin{equation} \label{eq:STFT}
X(k,m) = \sum_{n=0}^{N-1}w(n)x(n+mH)e^{-j2\pi nk/N} 
\end{equation}

\noindent
where $k$ is the discrete-frequency index, $m$ is the hop number for the analysis window, $N$ is the frame size (2048 samples), and $H$ is the hop size (512 samples). 
A Hanning window is used for the window function $w$, and the sampling frequency is \SI{44.1}{\kilo \hertz}.
$X(k,m)$ is evaluated for $k=0,...,N/2$ and the magnitude of these components are retained. 

% -----------------------------------
% Placed here so figure appears on same page of section
\begin{figure*}
    \centering
    \includegraphics[width=\linewidth]{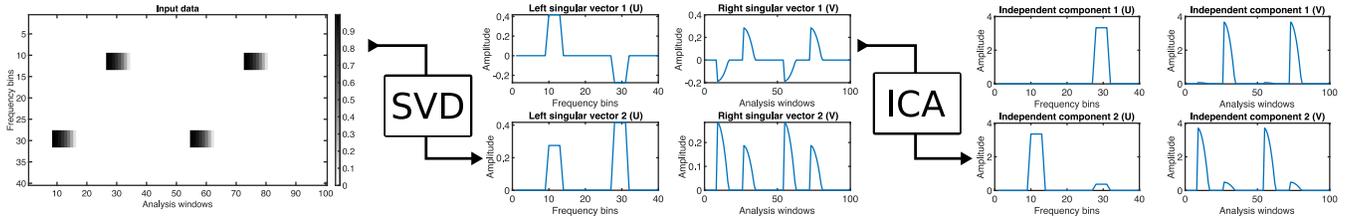}
    \caption{Illustration of the ISA algorithm on a sample magnitude spectrum. Input data (left) undergoes SVD to produce the frequency basis spectra $U$ and time-activation functions $V$. Applying ICA to the frequency-basis spectra and time-activation functions results in improved separation of the events, as can be seen in the four plots to right.}
    \label{fig:svd_overview}
\end{figure*}
% ----------------------------------------------

\subsection{Independent subspace analysis} \label{ssec:ISA}
% Implementation of ISA requires SVD followed by ICA to be carried out.
SVD first decomposes the input matrix $X$ into three matrices $U$, $V$, and $S$, to identify subspaces in order of the variance they contribute towards the original data \cite{uhle2003extraction}.
% Assuming the input matrix has dimensions $p \times q$,
% \begin{itemize}
%     \item $U$ are the left singular vectors with dimensions $p \times p$,
%     \item $V$ are the right singular vectors with dimensions $q \times q$,
%     \item $S$ are the singular values with dimensions $p \times q$, contained in the diagonal elements of $S$ arranged from largest to smallest.
% \end{itemize}
In the context of the magnitude spectrum $X$, where columns contain the magnitudes of frequency content for a single analysis window, $U$ are the frequency basis spectra and $V$ are the corresponding time-activation functions. 
The outer product of each spectra and time-activation pair produces a subspace of the original input data, and the input data can be reconstructed using
\begin{equation} \label{eq:SVD}
    X = USV^T
\end{equation}
For this algorithm, the columns of $V$ will act as time-activation functions for cough events in the input data $X$, and only the first nine singular values are computed (determined experimentally to be optimal).

% Each subsequent subspace produced by SVD contributes decreasing amounts of variance towards the input data.
\fref{fig:svd_overview} illustrates the input and output of SVD on an example magnitude spectrum.
These pairs of frequency-basis spectra and time-activation functions ($U_1, U_2, V_1, V_1 $) both contribute towards reconstructing elements of both ``sources'' in the magnitude spectrogram.
Ideally a single time-activation function would coincide with each single type of event has occurred, but in practice this is not the case.
This can be overcome by applying ICA to the time-activation function to transform the decorrelated time-activations into statistically independent activation functions \cite{hyvarinen2000independent,comon1994independent}.
% ----------------------------------------------
% \subsection{Independent component analysis} \label{ssec:ICA}
% SVD results in a set of decorrelated (but not independent) time-activation functions. 
This is the principle underlying ISA.
In \fref{fig:svd_overview}, the plots on the right show the output of the ICA stage.
The frequency-basis spectrum and time-activation function pair's contributions towards reconstructing both sources in the input magnitude spectrum has decreased significantly.

\subsection{Time-activation selection and thresholding}
Recall that in \sref{ssec:ISA} the first nine singular values were computed.
In determining which time-activation functions to retain, the kurtosis function $k(v)$ is used, 
\begin{equation} \label{eq:kurtosis}
    k(v) = \frac{\sum_{m=0}^{M-1}(V_{m,v} - \bar{V_{v}})^4 /M}{\sigma_{V_{v}}^{4}}
\end{equation}

\noindent
where $V_{m,v}$ is the $m^{th}$ sample in $v^{th}$ column of $V$, $\bar{V_{v}}$ and $\sigma_{V_v}$ are mean and standard deviation of $v^{th}$ column of $V$, respectively.
When normalised, kurtosis is 0 for a Gaussian distribution, positive for a ``peakier'' or leptokurtic distribution, and negative for a ``flatter'' or mesokurtic distribution.
Sparse events are likely to occur sporadically in the time-activation functions, meaning a majority of near-zero values, resulting in a leptokurtic distribution.
% This is the justification for using kurtosis to automatically identify suitable independent components that are more likely to represent coughs.
% In this algorithm, the time-activation functions with the 3 highest kurtosis measurements are summed and rectified to produce a candidate time-activation function $c$.
The time-activation functions with the three highest kurtosis measurements are retained and rectified to produce three candidate time-activation functions $c_1$, $c_2$, and $c_3$, for further analysis.

% ----------------------------------------------
% \subsection{Thresholding} \label{ssec:thresholding_and_summary}
% \subsection{Thresholding and summarisation} \label{ssec:thresholding_and_summary}
Peaks in each time-activation function correspond to moments in the input signal where the contribution of the corresponding frequency-basis spectrum is greater.
A threshold $\tau = a.\sigma_c$ is applied to the candidate time-activation functions to identify peaks,
% \begin{equation} \label{eq:threshold}
%     \tau = a.\sigma_c
% \end{equation}
% \noindent
where $\sigma_c$ is the standard deviation of $c$, and $a$ is a constant, with $4<a<8$. 
% in the range of 4 to 8, with 5 being used in this analysis. 

Retained peaks from the candidate time-activation function are used to generate a summary of the input signal, where \SI{1}{\second} windows around each peak are extracted from the input signal and concatenated to produce the short summary of the input signal.
% ,see \fref{fig:summary}. 
% \begin{figure}[hbtp!]
%     \centering
%     \includegraphics[width=\linewidth]{Figures/summary2.eps}
%     \caption{Example of input signal of \SI{600}{\second} and corresponding summary signal with a reduced duration of \SI{23}{\second}. The overall duration of the input signal is reduced significantly when the number of incorrect detections are maintained at a minimum.}
%     \label{fig:summary}
% \end{figure}

% ----------------------------------------------
\subsection{Dataset and evaluation} \label{ssec:dataset}
The dataset used in evaluating the proposed algorithm was constructed from a number of publicly available cough, non-cough, and background sources collected from YouTube videos \cite{kvapilova2019continuous} and the DCASE 2016 Challenge \cite{Benetos2016}.
% Scaper \cite{salamon2017scaper}, a soundscape synthesis and augmentation tool, was used to produce the synthesized test signals annotations.
Each test signal comprises 20 coughs produced by one individual and alternative foreground sounds (door knocks, table banging, speech, laughing, etc.).
In total, 10 test signals were created with a duration of 10 minutes each.
Instructions for reproducing these test signals, including annotations and URL links, can be accessed at \cite{leamy_2020_isacough}.  

% ----------------------------------------------
% \subsection{Evaluation}
The evaluation framework used here is adapted from \cite{bilen2019framework}, a framework for polyphonic sound-event detection which overcomes the limitations of collar-based event decisions and labelling subjectivity by annotators.
\textit{True positive} (TP) or \textit{false positive} (FP) decisions rely on the degree of overlap with annotated events. 

The overlap $t_o$ between an annotated event and detected event (see \fref{eq:detection_criteria}) is measured and expressed as a ratio of a \SI{500}{\milli \second} window, based on the average duration of coughs presented in \cite{leamy2019re}.
When this ratio exceeds the detection tolerance criteria $\rho_{DTC}$, the detected event is labelled a TP, otherwise FP is declared, 
\begin{equation} \label{eq:detection_criteria}
    D =
    \begin{cases}
      TP, & \text{if}\ \frac{t_o}{500ms} > \rho_{DTC}  \\
      FP, & \text{otherwise}
    \end{cases}
    \hspace{1mm} ,0 < \rho_{DTC} \leq 1
\end{equation}
where $D$ is a single detection and \mbox{$\rho_{DTC} =0.3$}.
Undetected annotated events are classed as \textit{false negatives} (FN).
\begin{figure}[htbp]
    \centering
    \includegraphics[width=\linewidth]{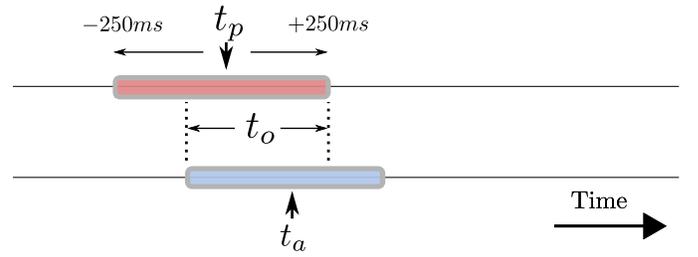}
    \caption{Overlap between the time of a detected event $t_p$ and time of an annotated event $t_a$ is used to determine if a detected event is classified as a TP or FP. The overlap time $t_o$ is expressed as a ratio of a \SI{500}{\milli \second} window. When this ratio exceeds the detection tolerance criteria $\rho_{DTC}$ the detected event is marked as a TP.}
    \label{fig:overlap_measurement}
\end{figure}

The performance of this algorithm is quantified using the true positive ratio ($r_{TP}$) and the false positive rate ($R_{FP}$) \cite{bilen2019framework},
\begin{equation} \label{eq:metrics}
    r_{TP} = \frac{N_{TP}}{P} \hspace{4mm} R_{FP} = \frac{N_{FP}}{T_{dur}}
\end{equation}
where $N_{TP}$ and $N_{FP}$ are the number of true and false positives, $P$ is the number of annotated cough events, and $T_{dur}$ is the duration of the signal being analysed.
% The length of the summarised signal is $T$.

% ----------------------------------------------
\section{Results and discussion} \label{sec:results}
The mean values for the true positive ratio, false positive rate, and summarised duration across all test signals are presented in \tref{tab:results_summary}.
These results were produced for each time-activation function. 
Activation function $c_1$ achieved the best performance in each metric across all mean values, suggesting this is the appropriate activation function to use.
The increase in false positive rate with $c_2$ and $c_3$ suggests that these activation functions encapsulate features common to both cough and non-cough events after SVD.  
\begin{table}[hbtp]
    \centering
    \caption{Summary of results from all candidate time-activation functions ($c_1$, $c_2$, and $c_3$) showing the mean results across all test signals. 
    % Activation function $c_1$ achieved the best performance across all metrics.
    }
    \begin{tabular}{lrrr}
        \\
        \toprule
        c & $r_{TP} (\%) $ & $R_{FP} (min^{-1})$ & $T (min)$\\
        \midrule
        1 & 76.00 & 2.85 & 0.73 \\ 
        2 & 67.00 & 4.17 & 0.92 \\ 
        3 & 39.00 & 6.26 & 1.17 \\  
        \bottomrule
    \end{tabular}
    \label{tab:results_summary}
\end{table}

\noindent
\tref{tab:results_best} highlights a best-case scenario which includes the metrics for each test signal and the associated activation function that achieved this result.
For the majority of results, $c_1$ produced the best case which aligns well with the results from \tref{tab:results_summary}.
For signal \#3 an \#5, the activation functions that achieved the most desirable results are $c_3$ and $c_2$, respectively.
Assuming that the appropriate time-activation functions are used, an improvement in the mean true positive ratio and false positive rate is observed. 
\begin{table}[hbtp]
    \centering
    \caption{A best-case scenario representation of the results, where $c$ is the $c^{th}$ activation function, $r_{TP}$ is the true positive ratio, $R_{FP}$ the false positive rate, and $T$ is the summarised signal duration.
    % An automatic method of determining the time-activation function that best represent the cough content will improved the results.
    }
    \begin{tabular}{lrrrr}
        \\
        \toprule
        \# & c & $r_{TP} (\%) $ & $R_{FP} (min^{-1})$ & $T (min)$\\
        \midrule
        1 & 1 & 70 & 4.5 & 0.98 \\
        2 & 1 & 90 & 3.4 & 0.87 \\
        3 & 3 & 95 & 1.4 & 0.55 \\
        4 & 1 & 100 & 4.00 & 1 \\
        5 & 2 & 100 & 0.6 & 0.43 \\
        6 & 1 & 90 & 2.2 & 0.67 \\
        7 & 1 & 95 & 0.5 & 0.4 \\
        8 & 1 & 90 & 6.7 & 1.42 \\
        9 & 1 & 95 & 3.6 & 0.92 \\
        10 & 1 & 100 & 1.2 & 0.53 \\
        \midrule
         \textbf{Mean} & & \textbf{92.50} & \textbf{2.81} & \textbf{0.77} \\
        \bottomrule
    \end{tabular}
    \label{tab:results_best}
\end{table}

From the results presented here, ISA has shown to be a reliable method for detecting the presence of cough sounds in audio recordings.
Referring to \tref{tab:results_best}, it can be seen that the majority of cough sounds are accurately represented within a single time-activation function, $c_1$ in this case, which supports the use of kurtosis as a suitable statistical measure for determining which activation function to use to achieve the most desirable results.
On average, kurtosis produces the best activation function to use as shown in \tref{tab:results_summary}. 
The mean results can be improved by selecting the time-activation function containing the coughs manually.
An automatic method of determining which activation functions could be used to complement this algorithm and improve overall results, as shown by the mean calculations in \tref{tab:results_best}.

% ----------------------------------------------
% \input{Sections/discussion}

% --------------------------------------------------------
\section{Conclusions}
% Good way to do this is to rephrase what the abstract and intro say
% With heath-care services currently under increased pressure due to the outbreak of COVID-19, 
% With COVID-19, a chronic cough is one of the symptoms used to diagnose those that have contracted the virus, and with early detection, isolation, treatment being the recommended approach around the world, an automatic method monitoring and tracking the progress of an infected individual's cough is paramount.
Building on an approach used for automatically transcribing drum sounds, the proposed algorithm uses ISA for detecting the presence of an individual's cough within audio recordings, reducing the human effort required by medical staff to identify cough events within long recordings. 
% Furthermore, this approach could be used to pre-process data sets which require human labelling for the purpose of training more general supervised learning algorithms. 
% For example, a human need only be presented with the summary recording for manual labelling. 

% Tested on synthetic ambulatory cough recordings,
An average true positive ratio of \SI{76}{\percent}, and a false positive rate of 2.85 false positives per minute was achieved, with a significant reduction in the duration of signals.
This algorithm's ability to detect cough sounds is evident, and a reduction in false positives will improve the overall performance.
% A solution to consider in future work utilises multiple analysis windows during the SVD stage, taking advantage of the time-dependant structure of a cough's evolution.
A solution to consider in future work utilises multiple analysis windows during the SVD stage, taking advantage of the varying spectral characteristics observed during different phases of a single cough event.
% Also, a comprehensive comparison to the state-of-the-art in cough detection to further determine the performance of this approach is to be conducted.
Also planned is a more detailed comparison against other state-of-the-art cough-detection approaches.

% As discussed in \sref{sec:results}, the time-activation functions produced by ISA have shown to be capable of detecting cough sounds in the synthesized audio recordings, with kurtosis also being a reliable metric for determining the most appropriate time-activation function that should be used on average.
% This algorithm would benefit from a method of analysing the time-activation functions in a manner that would further  reduce the false positive rate and subsequently reduce the duration of signals to be analysed by medical professionals. 

% References should be produced using the bibtex program from suitable
% BiBTeX files (here: strings, refs, manuals). The IEEEbib.bst bibliography
% style file from IEEE produces unsorted bibliography list.
% -------------------------------------------------------------------------
\bibliographystyle{IEEEtran}
\bibliography{mybib,cough_detection}

\end{document}